\documentclass[aps,prb,twocolumn,showpacs,superscriptaddress,floatfix,nofootinbib]{revtex4-1}
\usepackage{amsfonts,hyperref}

\usepackage{graphicx}
\usepackage{amsmath}
\usepackage{times}
\usepackage{amssymb}

\usepackage{hyperref}
\hypersetup{
        colorlinks=true,
        pdfauthor={Bela Bauer}
}

\newcommand{\be}{\begin{equation}}
\newcommand{\ee}{\end{equation}}
\newcommand{\bea}{\begin{eqnarray}}
\newcommand{\eea}{\end{eqnarray}}

\usepackage{xspace}
\newcommand{\sys}[2]{(#1+2) \times #2}
\newcommand{\su}[1]{SU($#1$)}

\newcommand{\oh}{$\frac{1}{2}$\xspace}
\newcommand{\eqnref}[1]{(\ref{#1})}

\begin{document}

\title{Three-sublattice order in the SU(3) Heisenberg model on the square and triangular lattice}
\author{Bela Bauer}
\affiliation{Theoretische Physik, ETH Zurich, 8093 Zurich, Switzerland}
\affiliation{Station Q, Microsoft Research, Santa Barbara, CA 93106}

\author{Philippe Corboz}
\affiliation{Theoretische Physik, ETH Zurich, 8093 Zurich, Switzerland}
\affiliation{Institute of Theoretical Physics, \'Ecole Polytechnique F\'ed\'erale de Lausanne (EPFL), CH-1015 Lausanne, Switzerland}



\author{Andreas M. L\"auchli}
\affiliation{Institut f\"ur Theoretische Physik, Universit\"at Innsbruck, A-6020 Innsbruck, Austria}
\affiliation{Max-Planck-Institut f\"{u}r Physik komplexer Systeme, N\"{o}thnitzer Stra{\ss}e 38, D-01187 Dresden, Germany}

\author{Laura Messio}
\affiliation{Institute of Theoretical Physics, \'Ecole Polytechnique F\'ed\'erale de Lausanne (EPFL), CH-1015 Lausanne, Switzerland}


\author{Karlo Penc}
\affiliation{Research Institute for Solid State Physics and Optics, H-1525 Budapest, P.O. Box 49, Hungary}

\author{Matthias Troyer}
\affiliation{Theoretische Physik, ETH Zurich, 8093 Zurich, Switzerland}

\author{Fr\'ed\'eric Mila}
\affiliation{Institute of Theoretical Physics, \'Ecole Polytechnique F\'ed\'erale de Lausanne (EPFL), CH-1015 Lausanne, Switzerland}


\date{\today}

\begin{abstract}
We present a numerical study of the SU(3) Heisenberg model of three-flavor fermions on the triangular and square lattice by means of the density-matrix renormalization group (DMRG) and infinite projected entangled-pair states (iPEPS). For the triangular lattice we confirm that the ground state has a three-sublattice order with a finite ordered moment which is compatible with the result from linear flavor wave theory (LFWT). The same type of order has recently been predicted also for the square lattice [PRL 105, 265301 (2010)] from LFWT and exact diagonalization. However, for this case the ordered moment cannot be computed based on LFWT due to divergent fluctuations. Our numerical study clearly supports this three-sublattice order, with an ordered moment of $m=0.2-0.4$ in the thermodynamic limit.
\end{abstract}


\maketitle

\section{Introduction}
Recent advances in experiments on cold atomic gases have raised interest in systems consisting of several flavors of interacting fermions, which can be realized for example as different hyperfine states of alkali atoms~\cite{wu2003,honerkamp2004} or nuclear spin states of ytterbium~\cite{cazalilla2009} or alkaline-earth atoms.~\cite{wu2003,gorshkov2010} A model Hamiltonian to describe such systems is the $N$-flavor fermionic Hubbard model given by
\begin{equation}
H = -t \sum_{\langle i,j \rangle} \sum_{\alpha} c_{i \alpha}^\dagger c_{j \alpha} + H.c. + U \sum_i \sum_{\alpha, \beta} n_{i \alpha} n_{i \beta},
\end{equation}
where $\alpha$, $\beta$ run over the different flavors, $\langle i,j \rangle$ runs over pairs of nearest neighbors on the lattice and $i$ runs over all sites of the lattice.

The two-flavor case corresponds to the spin-\oh Hubbard model. It is generally accepted that for sufficiently large $U$ and at half filling, i.e. when each lattice site is occupied by exactly one fermion, the ground state is an antiferromagnetic Mott insulator. In experiments on cold atoms, the transition to a Mott insulator has recently been observed;\cite{jordens2008,schneider2008} the observation of the antiferromagnetic spin order is still open. The spin-\oh Heisenberg model is believed to be a good low-energy model for the spin degrees of freedom.

For the more general case $N > 2$, it is expected that, at certain fillings, Mott insulating states will also emerge.\cite{gorelik2009,miyatake2010} However, the spin order (or flavor order) in this case is not understood yet. This has raised interest in generalizations of the spin-\oh Heisenberg model, namely \su{N} Heisenberg models. Analogous to the $N=2$ case, these are obtained as second-order expansion of the above $N$-flavor fermionic Hubbard model in $t/U$ at a filling such that each site is occupied by exactly one particle.  
The Hamiltonian is
\begin{equation} \label{eqn:su3}
H = J \sum_{\langle i,j \rangle} \sum_{\alpha, \beta} |\alpha_i \beta_j \rangle \langle \beta_i \alpha_j |
\end{equation}
where the first sum runs over pairs of nearest neighbors and the second sum over flavors. In this paper, we will focus on the case of the \su{3} Heisenberg model, where $\alpha, \beta \in \lbrace A,B,C \rbrace$. 
%

%

Note that this model is different from the \su{N} Heisenberg models studied in Refs.~\onlinecite{affleck1988-sun,marston1989,read1989,read1990,harada2003,kawashima2007,beach2009,hermele2009,hermele2011} where other irreducible representations of \su{N} have been considered, which can be labelled by different Young tableaus. The corresponding Young tableau of our model has one single box, i.e. the fundamental representation at each site. In Refs.~\onlinecite{affleck1988-sun, marston1989} the large-N limit with $N/2$ particles per site has been studied (a Young tableau with $N/2$ boxes in one column). 
In Refs.~\onlinecite{hermele2009,hermele2011} the large-N limit with representations with $m$ rows and $n_c$ columns  with fixed $N/m$ and $n_c$ are considered ($n_c=1$ in Ref.~\onlinecite{hermele2009}). Another possibility is to use conjugate representations on two sublattices,~\cite{marston1989,read1989,read1990,harada2003,kawashima2007,beach2009} which is accessible by Quantum Monte Carlo simulations without a sign problem,\cite{harada2003,kawashima2007,beach2009} in contrast to the  \su{3}  model considered in this paper.

\begin{figure}
  \centering
%
%
%
%
%
%
\includegraphics{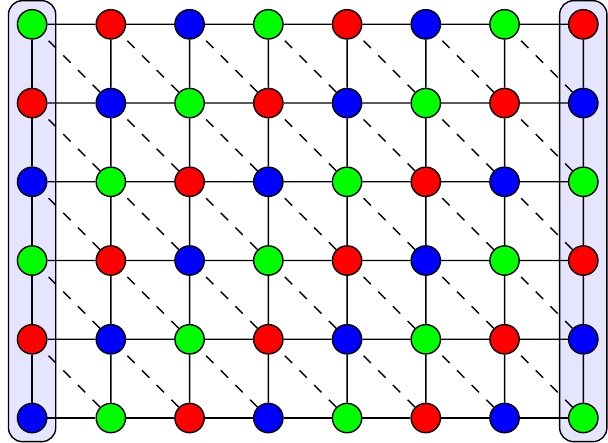}
  \caption{Proposed three-sublattice order for the \su{3} Heisenberg model on the square and triangular lattice. The triangular lattice is obtained from the square lattice by adding couplings along the dashed bonds shown above. Blue boxes indicate the sites that are pinned to a specific flavor in our DMRG simulations in order to explicitly break \su{3} symmetry. \label{fig:su3_order}}
\end{figure}
This \su{3} model is equivalent to the spin-1 bilinear-biquadratic model,
\begin{equation}
\label{eq:bb}
H = \sum_{\langle i,j \rangle} \left[ \cos{\theta} (\vec{S}_i \cdot \vec{S}_j )+ \sin\theta (\vec{S}_i \cdot \vec{S}_j)^2 \right].
\end{equation}
with $\theta=\pi/4$, thus when bilinear and biquadratic terms are equal and positive.

In a pioneering work, Papanicolaou~\cite{papa1988} studied the phase diagram of this model on the square lattice as a function of $\theta$ by a semiclassical analysis. Using a site-factorized variational ansatz (product state) he proposed that the case $\theta = \pi/4$ corresponds to a phase transition from the antiferromagnetically ordered phase (adiabatically connected to the purely bilinear case) to a "semi-ordered phase" with infinitely many degenerate ground states, including states with 2- or 3-sublattice order. For recent progress on the nature of the phases in the "semi-ordered" region and its vicinity, see Ref.~\onlinecite{toth2011}.


The situation changes for the closely related case of the triangular lattice (equivalent to introducing a coupling on one of the diagonals of each plaquette of the square lattice), where a site-factorized ansatz already predicts a three-sublattice order as depicted in Fig.~\ref{fig:su3_order}. \cite{tsunetsugu06, laeuchli2006-triang} This state is stable upon adding quantum fluctuations at the level of linear flavor wave theory (LFWT), and is supported by exact diagonalization results.\cite{laeuchli2006-triang} In contrast, on a zig-zag chain (the one-dimensional analog of the triangular lattice) the system undergoes spontaneous trimerization.~\cite{Corboz07}



A recent study based on LFWT indicates that on the square lattice a similar type of three-sublattice order is selected by quantum fluctuations.~\cite{toth2010} This type of order is further supported by exact diagonalization revealing a tower of states compatible with the continuous symmetry breaking of SU(3).~\cite{toth2010} The ordered moment of the symmetry broken state, however, cannot be computed at the level of LFWT because  fluctuations are divergent. Note that at this point this is an
artifact of the linear spin wave theory, and it is open whether higher order flavor wave corrections would lead to a finite or absent ordered moment. The only estimate of the ordered moment so far was obtained with exact diagonalization from the real-space correlation functions of an 18-site cluster, suggesting an ordered moment of  $60\%-70\%$ of the saturation value, which is expected to decrease with system size. To further establish the three-sublattice order on the square lattice it is important to have an estimate of the ordered moment in the thermodynamic limit.


While on both lattices three-sublattice order has been suggested, the mechanism how this order is selected is quite different: on the triangular lattice it is already favored at the classical level, and quantum fluctuations only renormalize the ordered moment, which is a situation similar to that of the \su{2} Heisenberg model on bipartite lattices. In the case of the square lattice, on the other hand, the three-sublattice order is one among the many degenerate states in the classical limit, which is selected by quantum fluctuations. Note that thermal fluctuations may select a different order.~\cite{toth2010}

In the previous semiclassical studies quantum fluctuations at the level of LFWT have been taken into account, and higher order terms have been neglected. Thus, it is still an open question if the three-sublattice order is stable upon including higher-order quantum fluctuations, or if in this case another state is selected.
An example of such a scenario has recently been observed in the SU(4) Heisenberg model on the square lattice,\cite{corboz11-su4} where low-order quantum fluctuations select a plaquette state, but additional higher-order quantum fluctuations finally favor a dimerized state. For the SU(3) model, exact results on small systems suggest that the order is stable,\cite{laeuchli2006-triang, toth2010} but an accurate numerical study for larger systems in two dimensions is so far missing.

In this paper we study the stability of the three-sublattice order of the model \eqref{eqn:su3} on the triangular and square lattice with state-of-the-art numerical simulations. We present results for finite 2D systems with open boundaries up to a size $8\times 8$ using the density matrix renormalization group (DMRG) method, and infinite 2D systems with infinite projected entangled-pair states (iPEPS). Both methods belong to the class of tensor network algorithms, enabling to compute ground state properties with an accuracy which can be systematically controlled by a refinement parameter, called the bond dimension.  Both methods confirm that the ground state has three-sublattice order for both type of lattices,
and we provide an estimate of the ordered moment in the thermodynamic limit. Finally we discuss an alternative approach based on Schwinger bosons. Unfortunately, this approach turns out to be unable to describe spontaneous SU(3) symmetry breaking, and, as a consequence, its results disagree with those of all other approaches regarding the type of ordering and the value of the ordered moment.

The outline of this paper is as follows. In Sec.~\ref{sec:methods} we give a short summary of linear flavor wave theory and provide details on the DMRG and iPEPS simulations. In Sec.~\ref{sec:triangular} we first present the results for the triangular lattice, where the three-sublattice order is expected to be more robust than on the square lattice, since this order is already favored at the classical level. We compare and discuss results for the energies and the ordered moment obtained with DMRG, iPEPS and LFWT.  In Sec.~\ref{sec:square} we provide a similar study for the square lattice case, where we find an ordered moment which is also finite, but stronger suppressed by quantum fluctuations than on the triangular lattice. Finally, Sec.~\ref{sec:summary} summarizes our results. In Appendix~\ref{sec:app} we report on our attempt to extend the Schwinger boson mean-field theory to SU(3).


\section{Methods}
\label{sec:methods}

\subsection{Linear flavor wave theory}
\label{sec:LFWT}

The linear flavor wave theory is the extension of the usual SU(2) spin wave theory to SU(N) models. It has been formulated in Refs.~\onlinecite{N1984281} and \onlinecite{papa1988} for the SU(3) case and in Ref.~\onlinecite{PhysRevB.60.6584} for the SU(4) case. For completeness, here we give some details for the three--sublattice order on the triangular and square lattice in the SU(3) Heisenberg model --- the cases under scrutiny in this paper. We note that for the triangular lattice, an analogous calculation has been presented by Tsunetsugu and Arikawa.\cite{tsunetsugu06}

We begin by extending the Hamitonian (\ref{eqn:su3}) to the case where on each site the states belong to the symmetrical irreducible representation of the SU(3) algebra that can be represented by Young-tableaux drawn with $M$ boxes arranged horizontally. The SU(3) spin operators in such a symmetrical irreducible representation can be expressed as
\begin{equation}
 S_{\beta}^{\alpha}(l) = b^{\dagger}_{\beta}(l) b^{\phantom{\dagger}}_{\alpha}(l) ,
\end{equation}
using Schwinger bosons with 3 flavors, where $l$ is the site index and the number of bosons on each site is
\begin{equation}
\sum_{\alpha\in\{A,B,C\}} b^{\dagger}_{\alpha}(l) b^{\phantom{\dagger}}_{\alpha}(l) =M,
\end{equation}
equal to the number of boxes in the Young tableau.
The $S_{\beta}^{\alpha}(l)$ operators satisfy  the SU(3) Lie algebra,
\begin{equation}
  \left[S_{\beta}^{\alpha},S_{\beta'}^{\alpha'} \right] =
  S_{\beta}^{\alpha'}\delta_{\beta'}^{\alpha} - S_{\beta'}^{\alpha}\delta_{\beta}^{\alpha'}
  \label{eq:Scommrel}
\end{equation}
 where $\delta_{\beta}^{\alpha}$ is the Kronecker $\delta$ function.
 For $M=1$ the
$S^\alpha_\beta(l)$ operators act on the 3-dimensional, fundamental
representation $|\alpha\rangle $ (where $\alpha=A, B,$ or $C$) of the
SU(3) algebra as $S^\alpha_\beta |\alpha\rangle =|\beta\rangle $ and $S^\alpha_\beta |\alpha'\rangle = 0$
if $\alpha' \neq \alpha$, with $i$ being the site index.

The Hamiltonian now can be written as
\begin{equation}
  {\cal H } = J \sum_{\langle i,j \rangle} S^\alpha_\beta(i) S^\beta_\alpha(j) \;,
  \label{eq:H_bosons}
\end{equation}
where the sum is over the nearest neighbor lattice sites, and over the repeated $\alpha$
and $\beta$ flavor indices. To draw a parallel to the SU(2) case, the Hamiltonian (\ref{eqn:su3}) in the fundamental irreducible representation corresponds to the spin--1/2 Heisenberg model, while the Hamiltonian (\ref{eq:H_bosons}) to the Heisenberg model of spins with length $S$ (actually for the SU(2) case $S=M/2$).

In the following, we consider an ordered state where the spins on the sites $l$, that belong to sublattice $\Lambda_\alpha$, point in the direction $\alpha$. Following the analogy with the spin wave theory that is a 1/S expansion, we take the $M \rightarrow \infty$ limit and do a $1/M$ expansion. Starting from the ordered state we can use the following expansion for the $S^\alpha_\beta(l)$  operators in the large--$M$ limit:
\begin{eqnarray}
  S^\alpha_\alpha(l) &=& M - \mu_\alpha(l), \\
  S^\alpha_\beta(l) &=& b^{\alpha\dagger}_{\beta}(l) \sqrt{M- \mu_\alpha(l)}
     \approx \sqrt{M} b^{\alpha\dagger}_{\beta}(l) ,
   \\
  S^\beta_\alpha(l) &=& \sqrt{M -  \mu_\alpha(l)} b^{\alpha}_{\beta}(l)
   \approx \sqrt{M} b^{\alpha}_{\beta}(l), \\
  S^{\beta'}_\beta(l) &=& b^{\alpha\dagger}_{\beta}(l) b^{\alpha}_{\beta'}(l),
\end{eqnarray}
where we have introduced the shorthand notation
\begin{equation}
  \mu_\alpha(l) = \sum_{\beta \neq \alpha} b^{\alpha\dagger}_{\beta}(l) b^{\alpha}_{\beta}(l) .
\end{equation}
The $b^{\alpha\dagger}_{\beta}(l)$ operators with $\beta\neq \alpha$ now correspond to the Holstein--Primakoff bosons on sublattice $\Lambda_\alpha$, and the superscript $\alpha$ keeps track of the sublattice. We replace the expressions above into Hamiltoniam~(\ref{eq:H_bosons}).
Expanding in $1/M$ and keeping the quadratic terms only, for the exchange
term between sites $l \in \Lambda_\alpha$ and  $l' \in \Lambda_{\alpha'}$ we get
\begin{eqnarray}
 \sum_{\beta,\gamma} S_{\beta}^{\gamma}(l) S_{\gamma}^{\beta}(l') &=& M
      \bigl[
         b^{\alpha\dagger}_{\alpha'}(l) b^{\alpha}_{\alpha'}(l)
       + b^{\alpha'\dagger}_{\alpha}(l') b^{\alpha'}_{\alpha}(l') \bigr. \nonumber\\
       &&
\bigl.       + b^{\alpha\dagger}_{\alpha'}(l) b^{\alpha'\dagger}_{\alpha}(l')
       + b^{\alpha}_{\alpha'}(l) b^{\alpha'}_{\alpha}(l')
      \bigr] .
\end{eqnarray}
in leading order in $M$ --- note that the bosons with flavor different from the ordered $\alpha$ and $\alpha'$ flavor are missing from the bond expression. Assuming a three-sublattice ordered state, we define the following
Fourier transformation:
\begin{equation}
  b^{\alpha}_{\beta,\mathbf{k}} = \sqrt{\frac{3}{N_\Lambda}} \sum_{l\in \Lambda_\alpha} b^{\alpha}_\beta(l)
  e^{i \mathbf{k} \mathbf{r}_l}
\end{equation}
where the summation is over the $N_\Lambda/3$ sites of the $\Lambda_\alpha$ sublattice ($N_\Lambda$ is the number of lattice sites).
The Hamiltonian between the sublattices $\Lambda_\alpha$ and $\Lambda_\beta$ in  $\mathbf{k}$--space reads
\begin{eqnarray}
  {\cal H }_{\alpha \beta} &=& \frac{z J M}{2}\sum_{k}
   \left[
          b^{\beta\dagger}_{\alpha,\mathbf{k}} b^{\beta}_{\alpha,\mathbf{k}} +
          b^{\alpha\dagger}_{\beta,-\mathbf{k}} b^{\alpha}_{\beta,-\mathbf{k}}
    \right.
 \nonumber\\
 &&  \left.
  + \gamma_\mathbf{k}
          b^{\alpha\dagger}_{\beta,-\mathbf{k}} b^{\beta\dagger}_{\alpha,\mathbf{k}}
    + \gamma^*_\mathbf{k}
          b^{\alpha}_{\beta,-\mathbf{k}} b^{\beta}_{\alpha,\mathbf{k}}  \right] ,
\end{eqnarray}
 where $z$ is the coordination number of the lattice ($z=4$ for the square and $z=6$ for the triangular lattice). The factor $\gamma_\mathbf{k}$ reads
\begin{equation}
 \gamma_\mathbf{k} = \frac{1}{3}
 \left(
 e^{i k_x} + 2 e^{-i k_x/2} \cos \frac{\sqrt{3}k_y}{2}
 \right)
\end{equation}
for the triangular lattice and
\begin{equation}
 \gamma_\mathbf{k} = \frac{1}{2}
 \left(
 e^{i k_x} + e^{i k_y}
 \right)
\end{equation}
for the square lattice, with  $\gamma^*_\mathbf{k} =  \gamma_{-\mathbf{k}} $.

The full Hamiltonian is $\mathcal{H} = \sum_{\alpha<\beta} \mathcal{H}_{\alpha\beta}$.
It can be diagonalized via a Bogoljubov transformation:
\begin{equation}
  \left(
    \begin{array}{c}
      \tilde b_{\alpha,\mathbf{k}}^{\beta\dagger} \\
      \tilde b_{\beta,-\mathbf{k}}^{\alpha}
    \end{array}
  \right)
  =
  \left(
    \begin{array}{cc}
   \cosh \theta(\mathbf{k}) &
   \sinh \theta(\mathbf{k}) \\
   \sinh \theta(\mathbf{k}) &
   \cosh \theta(\mathbf{k})
    \end{array}
  \right)
  \left(
    \begin{array}{c}
      b_{\alpha,\mathbf{k}}^{\beta\dagger} \\
      b_{\beta,-\mathbf{k}}^{\alpha}
    \end{array}
  \right)
\end{equation}
with
$ \tanh 2 \theta(\mathbf{k}) = \gamma_\mathbf{k}$ ,
leading to
\begin{equation}
 \mathcal{H} = - \frac{z}{2} J M N_{\Lambda}
 + M \sum_{\mathbf{k}\in \textrm{RBZ}} \sum_{\alpha} \sum_{\beta\neq\alpha}
  \omega(\mathbf{k})
 \left[
  \tilde b^{\alpha\dagger}_{\beta,\mathbf{k}} \tilde b^{\alpha}_{\beta,\mathbf{k}} +\frac{1}{2}
 \right]
 \;.
\end{equation}
The dispersion of the flavor waves is given by
\begin{equation}
  \omega(\mathbf{k})
= \frac{z}{2}J \sqrt{ 1 - |\gamma_\mathbf{k}|^2}
\end{equation}
There are 6 degenerate branches in the reduced Brillouin zone, which is equivalent to 2 branches in the normal Brillouin zone. The dispersion agrees with the result of Tsunetsugu and Arikawa\cite{tsunetsugu06} for the triangular lattice. For the square lattice, it is given in Ref.\onlinecite{toth2010}.

The energy per site due to quantum fluctuations is given by the expression
\begin{eqnarray}
 \left( 2 \left\langle \frac{\omega(\mathbf{k})}{2} \right\rangle_{BZ} -\frac{z}{2}J\right) M \label{eq:Ezp},
\end{eqnarray}
where we take into account that there are two modes per lattice site. The $\left\langle \dots \right\rangle_{BZ}$ denotes the average over the Brillouin zone.
Quantum fluctuations lower the energy  from 0 to $-0.630 J$ per site for the triangular and to $-0.727J$ for the square lattice.
Note that the energy per site of the triangular lattice is higher than the one of the square lattice despite the larger coordination
number of the former lattice.

The reduction of the ordered moment is calculated as
\begin{equation}
  \langle S^\alpha_\alpha(l) \rangle =
    M - \langle \mu_{\alpha}(l) \rangle
 =  M - \left\langle \frac{1}{\sqrt{ 1 - |\gamma_\mathbf{k}|^2}}-1 \right\rangle_{\rm BZ}.
\end{equation}
In the triangular lattice $\langle S^\alpha_\alpha(l) \rangle =  M - 0.516$, so that the on--site moment is reduced from 1 to $0.484$. In the square lattice, the reduced moment diverges due to the zero line in the spectrum. Thus, LFWT is unable to make a prediction for the ordered moment. We have tried to use a Schwinger boson mean-field theory (SBMFT)
to restore a gap along this line and remove the divergence (see Appendix \ref{sec:app}).
Unfortunately, the SBMFT turned out to be unsatisfactory in several respects, and its results regarding the ordered moment are not reliable.

\subsection{DMRG for finite two-dimensional systems}
\subsubsection{Setup}

For our DMRG simulations, we map the two-dimensional system to a chain following a "TV screen" method (sweeping along the vertical direction). We will generally refer to the extent in the horizontal direction as length, and in the vertical direction as width of the system. We use a single-site optimization scheme augmented by the improvement suggested in Ref.~\onlinecite{white2005}. We perform the simulation starting from different initial states and increase the bond dimension very quickly with the number of sweeps to avoid getting trapped in local minima. This is particularly important for the case of the square lattice, where an insufficient bond dimension may lead to unphysical states. Together with the large number of operators, this limits the bond dimension that we can reach with our computational resources to about $D \sim 5000$ states. Due to the huge growth of entanglement with the width of the system, this allows us to obtain sufficient accuracy for systems up to width 8.

\subsubsection{Calculation of the order parameter}

We expect that, in the thermodynamic limit, the \su{3} symmetry is spontaneously broken. If an appropriate basis is chosen (i.e. after an appropriate \su{3} rotation), one flavor becomes stronger on each site, i.e.
\begin{equation}
n_\alpha > n_\beta = n_\gamma \ \ \ \ \ \alpha,\beta,\gamma \in \lbrace A,B,C \rbrace.
\end{equation}
In this case, we can define the local moment
\begin{equation} \label{eqn:moment}
\langle m \rangle = \frac{3}{2} \left( \max_{\alpha = A,B,C} \langle n_\alpha \rangle - \frac{1}{3} \right),
\end{equation}
which should acquire a finite value in the range $\langle m \rangle \in [0,1]$.

On finite systems, the symmetry is never broken spontaneously and one would conventionally use the relation
\begin{equation}
\langle n_\alpha \rangle^2 =\lim_{d \rightarrow \infty} \left( \langle n_{\alpha,i} n_{\alpha,i+3d} \rangle -  \langle n_{\alpha,i} \rangle\langle n_{\alpha,i+3d} \rangle \right)
\end{equation}
to extract information about the moments. This however requires large systems and very accurate estimates for the correlation functions, which are hard to obtain from a DMRG simulation in two dimensions. We therefore follow the prescription of Refs.~\onlinecite{white2007, stoudenmire2011} and break \su{3} symmetry explicitly by introducing fields at the boundaries of the system. The local moments can then be measured locally, preferably on sites far away from the pinning fields. The pinning fields also fix the direction of the symmetry breaking to be along the basis vectors.

We introduce a column of pinned sites at each end of the system, as shown in Fig.~\ref{fig:su3_order}. We choose the system sizes such that the unpinned sites form a square, i.e. the system size including pinned sites is $\sys{L}{L}$. Pinning is done with a flavor-specific chemical potential of magnitude 1. In addition, such pinning fields reduce the entanglement in the system. Simulations were performed for both open and cylindrical boundary conditions.


\subsubsection{Boundary conditions}

An important question when performing finite-size DMRG simulations is the appropriate choice of boundary conditions. From an entanglement point of view, open boundary conditions appear favorable; also, these will have fewer long-range operators in the mapping to a chain. From a physical point of view, on the other hand, periodic boundary conditions are often preferred as they eliminate boundary effects. A compromise suggested e.g. in Ref.~\onlinecite{white2007} is to use cylindrical boundary conditions, which are favorable from an entanglement point of view.


Physically, such boundary conditions are compatible with the approach of pinning two columns, which preserves translational invariance in the vertical direction. In order not to frustrate the three-sublattice order, such boundary conditions should only be chosen for systems whose width is a multiple of three. For other system sizes, shifted cylindrical boundary conditions can be used. For example, for a system of width 5, the bottom site of column $i$ must be connected to the top site of column $i+1$ to obtain a system without additional frustration.

We find numerically that cylindrical boundary conditions favor a state that is a product of periodic length-6 chains wrapped around the cylinder. A calculation for the same model on a periodic chain of length 6 shows that the energy per site is very low in this case, making it favorable for small clusters. Such a state shows significantly reduced local moments. By choosing open boundary conditions in all directions, we suppress this effect.

Another subtlety occurs for the square lattice system sizes $L=3 n-1$, with $n$ a positive integer number, for which the pattern of our pinning fields allows two different ordered states, corresponding to the two different orientation of the diagonal
stripes.\cite{toth2010} For these cases, at a sufficiently large bond dimension, a superposition of both types of order will occur and lead to a significant decrease of the local moment (except on sites where the two types of order coincide) and a significant increase of the entropy. In these cases, we pin two additional sites to uniquely select the order.

\subsubsection{Extrapolation}

The reliable extrapolation of results obtained for a limited bond dimension to the limit of infinite bond dimension, where DMRG becomes exact, remains a challenge. While some reliable results are known for one-dimensional critical systems,\cite{tagliacozzo2008,pollmann2009} one has to resort to heuristic techniques in more general situations, such as two-dimensional systems. Such techniques include the extrapolation in the truncated weight~\cite{white2007} or in the variance. Since we use a single-site optimization method, the truncated weight cannot be obtained reliably, and the calculation of the variance is only possible for smaller system sizes. We therefore resort to an extrapolation of the magnetization in the bond dimension using the values obtained for the three largest values of $D$; for bond energies, we use only the result obtained for the largest value of $D$. While most simulations were performed using up to $D=4800$ states, we have confirmed the accuracy of our results with up to $D=6400$ states for some selected systems.

Similarly, an accurate finite-size extrapolation is difficult given the few system sizes we can access. Also, the dependence of the order parameter and the energy on the system size, boundary conditions and aspect ratio is not known. In fact, previous studies have even observed surprising cases such as non-monotonic behavior for very small systems.~\cite{white2007}


\subsection{Infinite projected entangled-pair states (iPEPS)}
\subsubsection{Setup}
An iPEPS is a tensor network made of a set of rank-5 tensors periodically repeated on a two-dimensional lattice to efficiently represent ground state wave functions in the thermodynamic limit. \cite{sierra1998, nishino1998, verstraete2004, nishio2004, murg2007, jordan2008} Each tensor has four auxiliary bonds which connect to the four nearest-neighbor tensors, and a fifth index carrying the local Hilbert space of a lattice site. The accuracy of the ansatz can be controlled by the dimension of the auxiliary bonds, called the bond dimension $D$. As the optimization scheme for the tensors we perform an imaginary time evolution with the so-called simple update (see Refs.~\onlinecite{jiang2008,corboz2010}) adopted from the time-evolving block decimation method in one dimension.~\cite{vidal2003-1, orus2008} For the square lattice we verified the results up to $D=8$ also with the full update (see e.g.~\onlinecite{corboz2010}), which is optimal but has a higher computational cost. The triangular lattice simulations are done with the same ansatz as for the square lattice, but now with an additional next-nearest neighbor interaction along one of the diagonal directions. The update scheme for this case is explained in Ref.~\onlinecite{corboz2010-nn}.

We performed simulations with different rectangular unit cells of size $L_x \times L_y$ in iPEPS.\cite{corboz2011} To represent the state with three-sublattice order efficiently, a $3 \times 3$ cell is used, with 3 different tensors $T_A, T_B,$ and $T_C$ for the three sublattices respectively. We verified that the same state is obtained by using a similar cell with 9 different tensors. The $2\times 2$ unit cell is used to enforce a state with two-sublattice order.

To contract the tensor network efficiently, e.g. for the computation of observables, the corner transfer matrix scheme\cite{orus2009-1} adapted to large unit cells\cite{corboz2010-nn} is used. The accuracy of the approximate contraction can be controlled by the so-called boundary dimension $\chi$. For large values of $D$ a $\chi$ up to 250 is used, where quantities of interest are extrapolated in $\chi$, with an extrapolation error being small compared to symbol sizes. For a better efficiency we use tensors with $\mathbb{Z}_q$ symmetry, a discrete abelian subgroup of SU(3).\cite{cincio2008,singh2010-1,bauer2011}

\subsubsection{Calculation of the order parameter}
Since iPEPS is an ansatz for the wave function in the thermodynamic limit, the SU(3) symmetry may be spontaneously broken, leading to a finite local moment $m$ defined in Eq.~\ref{eqn:moment}. In order to pin the direction of the moment in SU(3) color space an initial field is applied, which is taken to zero at a later stage of the imaginary time evolution. We verified that we obtain the same results without initial field, and by computing the moment taking all generators of SU(3) into account (see Eq.~\eqref{eq:m} in appendix \ref{sec:app}).


\subsubsection{Extrapolation}
\label{sec:extrapD}
For highly entangled systems quantities of interest such as the energy or the local moment are typically not converged as a function of the bond dimension $D$ at the maximal value of $D$ used, and thus an extrapolation to the infinite $D$ limit is desirable. However, in general the dependence of observables on $D$ is (still) unknown, which limits the accuracy of such extrapolations. 
Since the approach is variational the energy decreases with increasing $D$, and therefore the energy at the largest value of $D$ provides an upper bound of the exact energy. Empirically, the exact value lies between the linear extrapolated value and the value at the largest $D$, and thus we take the middle of these two values as an estimate and the difference between the two values as an error bar. 
The same holds for the local moment, which is typically suppressed with increasing $D$, since more quantum fluctuations are taken into account with increasing $D$ which renormalize the ordered moment. Typically, the energy converges faster than the local moment.

%


\section{Results for the triangular lattice}
\label{sec:triangular}
We first present the results for the energy and the ordered moment for the model on the triangular lattice, obtained with LFWT, ED (energies only), DMRG and iPEPS. As mentioned in the introduction, on the triangular lattice a three sub-lattice order is already obtained from a simple product state ansatz. Inclusion of quantum fluctuations via LFWT does not destroy the order,\cite{laeuchli2006-triang} but renormalizes the local moment. In the following we show that this holds even when including further quantum fluctuations.


\begin{figure}
  \centering
  \includegraphics[width=\columnwidth]{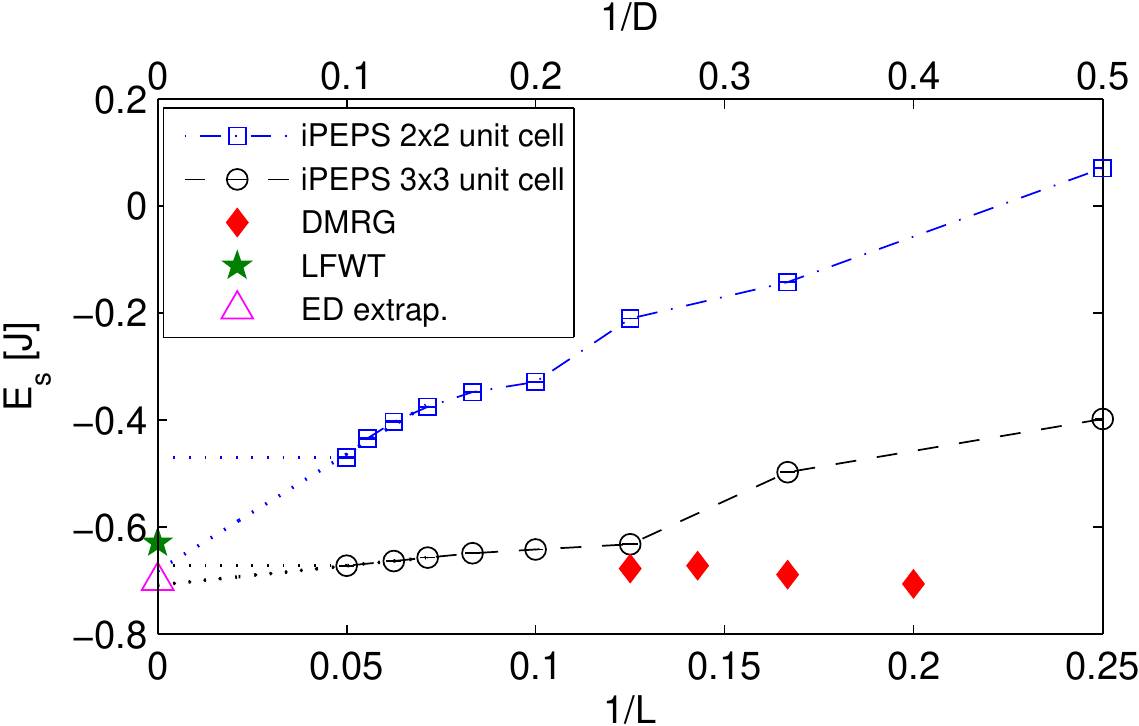}\\ \vspace{0.2cm}
  \includegraphics[width=\columnwidth]{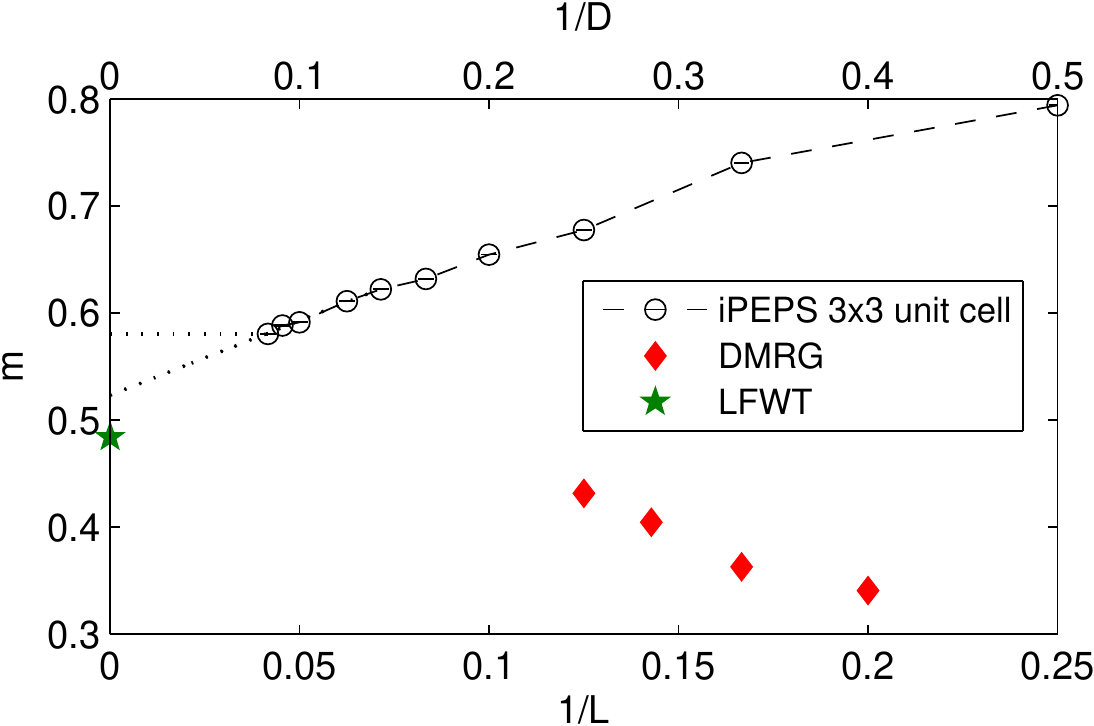}
  \caption{Comparison of the energy per site (upper panel) and the local moment (lower panel) of the SU(3) model \eqref{eqn:su3} on the triangular lattice, obtained from LFWT, ED, DMRG and iPEPS. In each plot, the DMRG results are shown as a function of the inverse system length $1/L$ (lower x-axis), whereas the iPEPS results are shown as a function of inverse bond dimension $1/D$ (upper x-axis). Dotted lines are only guides to the eye.}
  \label{fig:t_em}
\end{figure}

\subsection{Energy per site}
Figure~\ref{fig:t_em}a) shows a comparison of the energy per site obtained with the four methods, where linear flavor-wave theory predicts a value of $E_s=-0.6295 J$. Extrapolating
ED energies for symmetric clusters consisting of $N= 9$, $12$, and $21$ sites using a standard $1/N^{3/2}$ form, we obtain $E_s \approx -0.69(1) J$.%

\begin{figure}
  \centering
  \includegraphics[width=7cm]{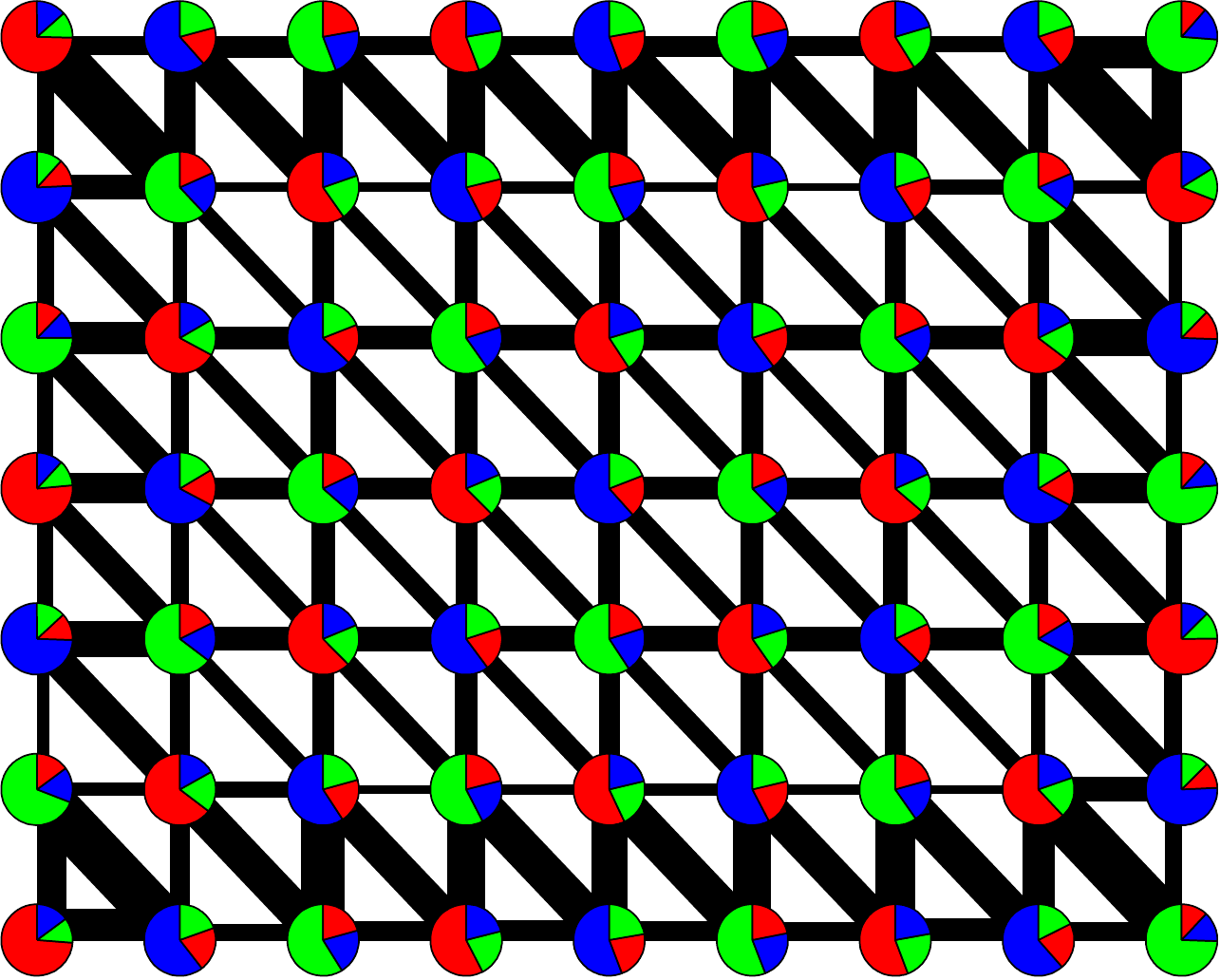}
  \caption{Bond energies and local color densities in the triangular $(7+2)\times7$ lattice obtained from DMRG. The thickness of the bonds is proportional to the magnitude of the bond energy. An external potential is applied on the first and the last column to pin the sites to a specific color.
  }
   \label{fig:conf_t}
\end{figure}

Since we use open boundary conditions in the DMRG simulations, the energy per site is not uniform in the system. Figure~\ref{fig:conf_t} shows that the energy per bond close to the boundary is lower than far away from the boundaries. To obtain an estimate of the energy per site in the "bulk" we average over the six bond energies around the central site for odd systems sizes. For even system sizes, we average over the four sites at the center of the system. This estimate is plotted in Fig.~\ref{fig:t_em}a) for different system sizes. The energy first increases with system size, and decreases slightly from $L=7$ to $L=8$. For the largest system $L=8$ the estimated energy per site in the bulk is $E_s=-0.6775J$.

Comparing the iPEPS energies obtained with different unit cell sizes, we find that the $3\times3$ unit cell yields a considerably lower variational energy than the $2\times2$ unit cell, which indicates that the symmetry breaking in the ground state is compatible with the 3-sublattice order. The energy per site has not converged yet as a function of bond dimension $D$. Since the energy typically converges faster than linearly in $1/D$ we (empirically) expect the energy to lie in between the value for the largest $D$, $E_s^{D=10}=-0.672 J$, and the energy obtained from linear extrapolation of the last three data points in $1/D$, $E_s^{\text{ex}}=-0.708 J$. Taking the mean of these two values yields an estimate of $E_s=-0.69(2)$, which is compatible with the DMRG result for the largest system.


\subsection{Local moment}
In Fig.~\ref{fig:t_em}b) we present the results for the local moment $m$ obtained with the various approaches, where $m=1$ for the fully polarized case.

As mentioned in Sec.~\ref{sec:LFWT}, linear flavor-wave theory predicts a value of $m=0.484$.

\begin{figure}
  \centering
  \includegraphics[width=\columnwidth]{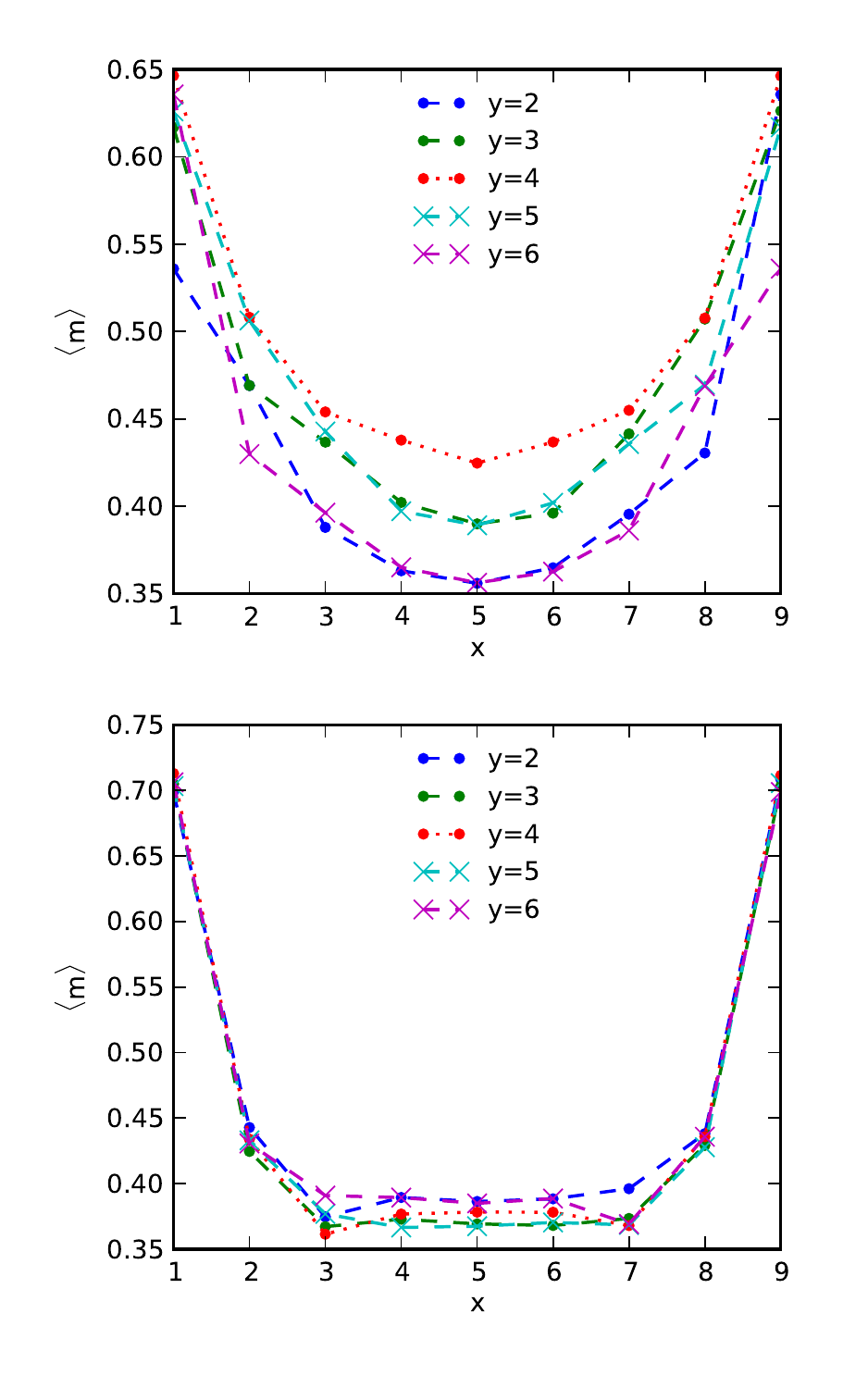}
  \caption{DMRG results with $D=4800$ states: Local moments as defined in Eqn.~\eqnref{eqn:moment} for the triangular (top panel) and square (bottom panel) lattice for a system size $\sys{7}{7}$. The plateau is very flat in the case of the square lattice, while corrections from the boundary are more pronounced for the triangular lattice. \label{fig:plateaux} }
  \label{fig:dmrg_mi}
\end{figure}

The DMRG results correspond to the local moment at the central site of the system for odd system sizes. For even system sizes, the magnitude of the ordered moment is averaged over the four sites that make up the central plaquette of the system. Variations depending on the distance to the boundaries in x- and y- direction can be observed, as shown in Fig.~\ref{fig:dmrg_mi}a). The value is decreasing with increasing distance from the pinning sites (in x-direction), whereas the value is seen to increase away from the boundary in y-direction.  As a function of system size the local moment is increasing. As mentioned before, an accurate extrapolation to the thermodynamic limit is challenging, but a value in the range $m \approx 0.43-0.6$ seems compatible with the DMRG data.

The local moment obtained with iPEPS decreases with increasing $D$, an effect which can also be observed e.g. in the SU(2) Heisenberg model. With increasing $D$ more quantum fluctuations are taken into account which reduce the magnetic moment from its value in the classical (product-state) limit, corresponding to $D=1$.  For the largest bond dimension used, $D=12$, we find a value $m=0.58$, whereas a linear extrapolation in $1/D$ suggests a value of $m=0.52$. As discussed in Sec.~\ref{sec:extrapD} the exact scaling behavior of $m$ with $1/D$ is not known, but empirically, we expect $m$ to lie in between these two values.

\subsection{Discussion}
Even though we cannot determine $m$ up to a high precision all three methods clearly suggest that the ground state has three-sublattice order with a large local magnetic moment in the range $m= 0.43 - 0.6$. As in the case of the SU(2) Heisenberg model on the square lattice, linear flavor wave theory (spin wave theory) already gives a good estimate of the local moment.

\label{sec:square}

\section{Results for the square lattice}
We next consider the SU(3) Heisenberg model on the square lattice. As explained in the introduction a site factorized ansatz leads to an infinite number of degenerate ground states and quantum fluctuations (with LFWT) selects the three-sublattice state.\cite{toth2010} Thus, quantum fluctuations seem to play a more dominant role on the square lattice, and it is conceivable that another ground state is selected when further quantum fluctuations beyond LFWT are taken into account. However, we show in the following that this is not the case here, i.e. that the three-sublattice order is stable and that additional quantum fluctuations only further renormalize the local moment.


\begin{figure}
  \centering
  \includegraphics[width=\columnwidth]{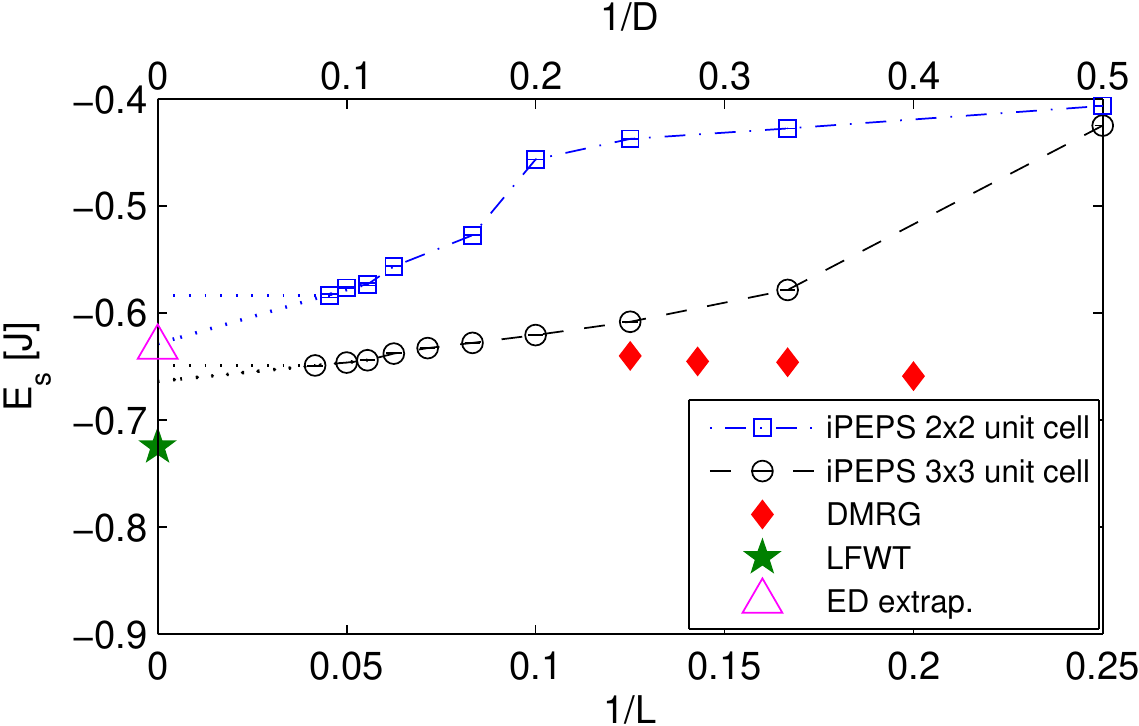}\\ \vspace{0.2cm}
    \includegraphics[width=\columnwidth]{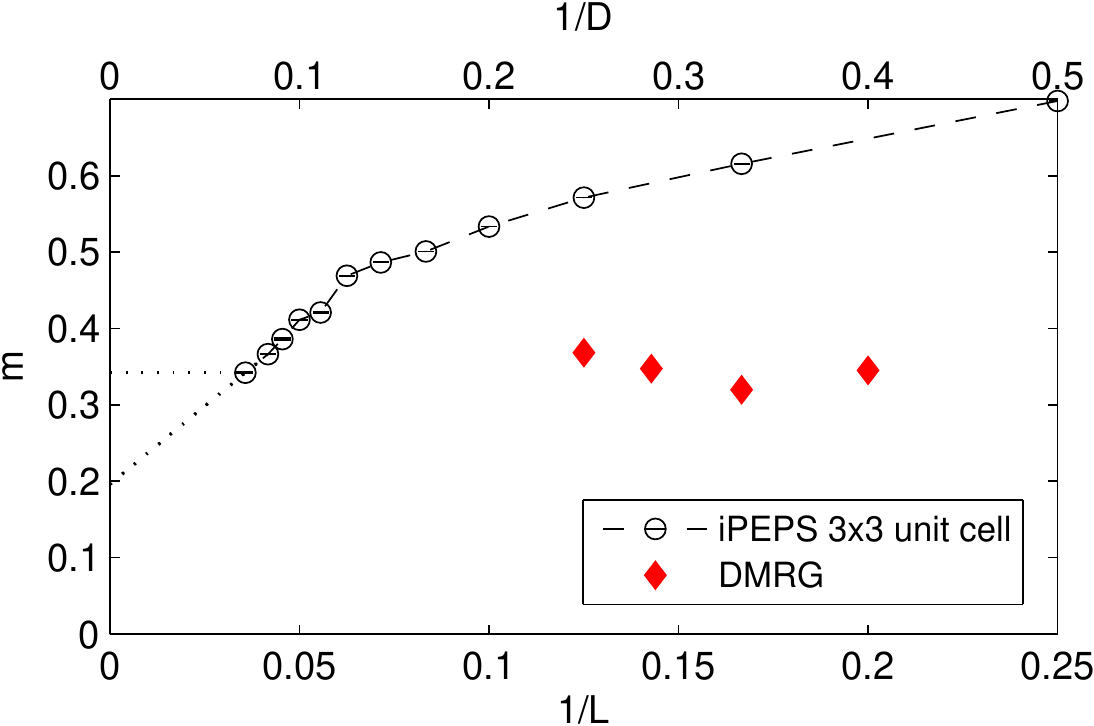}
  \caption{Same plot as in Fig.~\ref{fig:t_em} but  for the SU(3) model \eqref{eqn:su3} on the square lattice.}
    \label{fig:s_em}
\end{figure}

\subsection{Energy per site}
In Fig.~\ref{fig:s_em}a), the value of the energy per site from linear flavor-wave theory, $E_s=-0.725J$, has previously been calculated in Ref.~\onlinecite{toth2010}, and is low compared to the numerical results. We note the LFWT energy is not variational, so that it can be lower than the exact ground state value. We have also included an ED estimate of the energy per site $E_s = -0.63185 J$, which is based on an extrapolation using square samples with $N=9$ and $18$ sites.

\begin{figure}
  \centering
  \includegraphics[width=7cm]{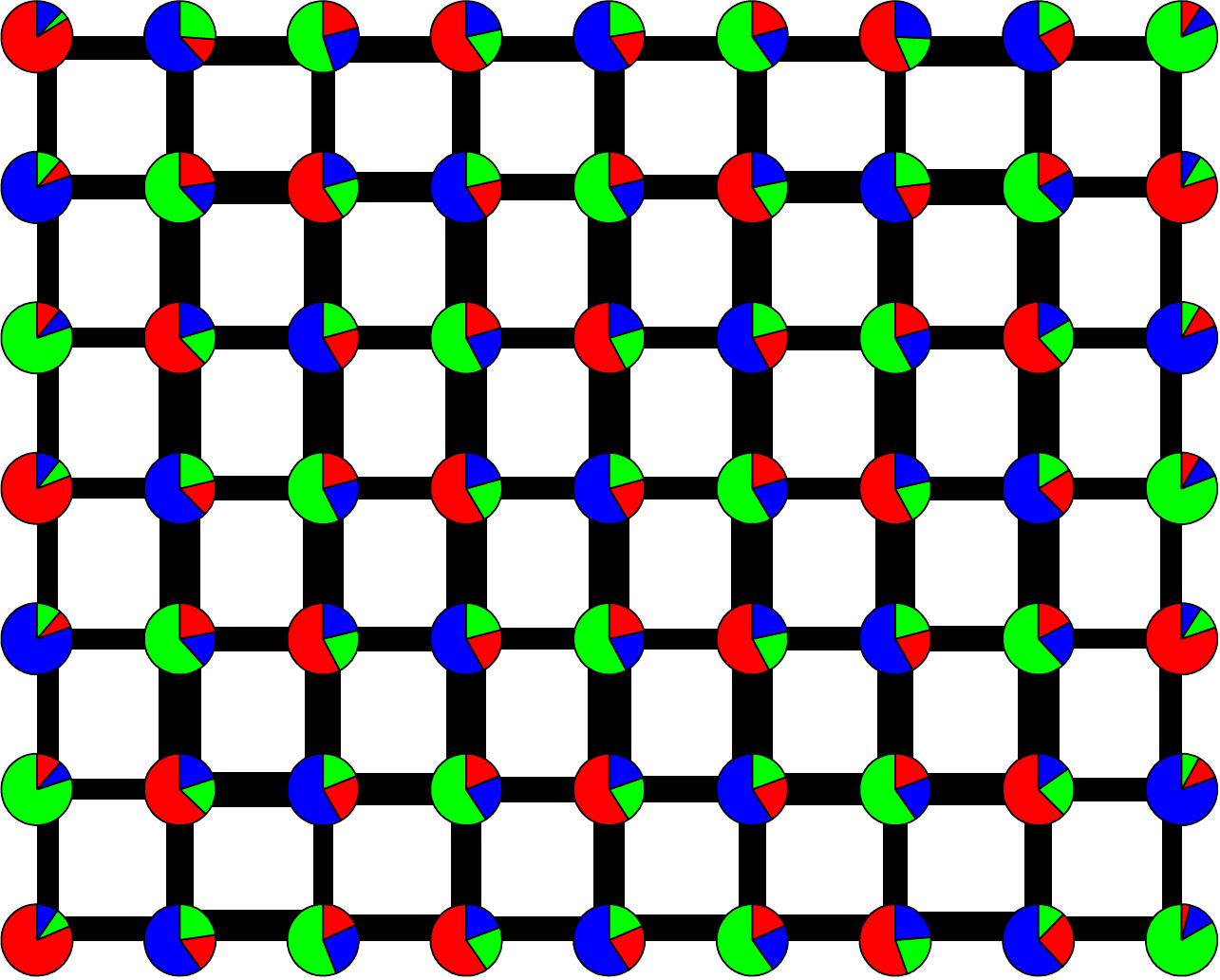}
  \caption{Bond energies and local color densities in the square $(7+2)\times7$ lattice obtained from DMRG. The thickness of the bonds is proportional to the magnitude of the bond energy. An external potential is applied on the first and the last column to pin the sites to a specific color.}
   \label{fig:conf_sq}
\end{figure}

The DMRG energy per site is $E_s=-0.625$ for the largest system, and seems to further increase as a function of system size. As in the triangular lattice case we estimate the bulk energy by taking the mean value over the bonds adjacent to a central site for odd system sizes and four sites for even system sizes. This energy seems higher than the LFWT and iPEPS result, which could indicate that boundary effects are large so that we do not get a good estimate for the "bulk" energy, or it could be that for larger systems the energy as a function of system size decreases again. We further note that an anisotropy in the bond energies can be observed, with stronger bonds in y-direction than in x-direction, shown in Fig.~\ref{fig:conf_sq}.

Comparing the energies from different unit cell sizes in iPEPS, we observe a similar behavior as on the triangular lattice, namely that the $3\times3$ unit cell provides a better variational energy than the $2\times2$ unit cell  for all values of $D$. The estimated energy per site in the limit $D\rightarrow \infty$ is $E_s=-0.66(1)$.

\subsection{Local moment}
Figure~\ref{fig:s_em}b) summarizes our results for the local moment, obtained from DMRG and iPEPS. As explained in Ref.~\onlinecite{toth2010}, the ordered moment cannot be computed within LFWT.

The finite size effects observed in DMRG are qualitatively different from the triangular lattice case.  The local moment as a function of distance of $x$ in Fig.~\ref{fig:dmrg_mi} reaches a plateau already after 3 sites away from the border, which could suggest that finite size effects on the ordered moment are smaller than on the triangular lattice. In Fig.~\ref{fig:s_em}b) the local moment in the middle of the system first decreases and then increases with system size with, however, a smaller slope than in the triangular lattice case. Thus, the present data is compatible with a non-vanishing local moment in the thermodynamic limit, which is smaller than on the triangular lattice. The ordered moment of the largest system is $m= 0.368$.

The local moment obtained with iPEPS decreases with increasing bond dimension but is not seen to extrapolate to zero in the limit $D \rightarrow \infty$. The value for the largest bond dimension, $D=16$, is $m=0.3422 $ which is close to the DMRG result for the largest system. However, in the limit $D\rightarrow \infty$ the data suggests a lower value of roughly $m=0.25(5)$, which is lower than the prediction from DMRG.

\subsection{Discussion}
Both the DMRG and iPEPS results are compatible with the proposed 3-sublattice N\'eel ordered ground state. From the present data we can only give a rough estimate of the ordered moment in the thermodynamic limit of $m=0.2-0.4$, which is clearly finite, but smaller than on the triangular lattice.



\section{Summary}
\label{sec:summary}
Our study confirms that the ground state of the SU(3) Heisenberg model exhibits a three-sublattice order on both the triangular and the square lattice, in accordance with previous predictions by LFWT and exact diagonalization.  \cite{tsunetsugu06, laeuchli2006-triang,toth2010}

The situation on the triangular lattice resembles the one of the SU(2) Heisenberg model on the square lattice. In both cases the ground state can already be understood at the classical level, and quantum fluctuations simply renormalize the ordered moment. These fluctuations are well captured already within linear flavor wave theory (i.e. spin wave theory in the SU(2) case). With iPEPS the ordered moment decreases with increasing bond dimension $D$, which can intuitively be understood because the bond dimension controls the amount of quantum fluctuations taken into account. All three methods used in this study yield a finite ordered moment in the range $m=0.43-0.6$. The uncertainty in this value stems from the error in the extrapolation to the thermodynamic limit in the case of DMRG, and from the extrapolation to the infinite $D$ limit in the case of iPEPS.

In the case of the square lattice, the order cannot be predicted at the classical level. Quantum fluctuations hence play a very different role than in the case of the triangular lattice: instead of renormalizing the mean-field result, they stabilize the three-sublattice order against other competing states. Quantum effects are therefore more important both qualitatively and quantitatively, and an estimate of the ordered moment in the thermodynamic limit has previously been lacking. Both DMRG and iPEPS predict a finite value in the range $m=0.2-0.4$, i.e. the ordered moment is more strongly suppressed than on the triangular lattice, but clearly finite.

\acknowledgments

We acknowledge helpful discussions with S. Manmana and U. Schollw\"ock, and the financial support of the Swiss National Fund and of MaNEP, and of the Hungarian OTKA Grant No. K73455. The DMRG code was developed with support from the Swiss platform for High-Performance and High-Productivity Computing (HP2C)~\cite{hp2c} and based on ALPS libraries.~\cite{ALPS_2,bauer2011-alps} Simulations were performed on the Brutus cluster at ETH Zurich.

\appendix

\section{Schwinger boson study}
\label{sec:app}

The Schwinger boson mean-field theory (SBMFT), introduced by Arovas and Auerbach\cite{Auerbach_largeN} and extended by Read and Sachdev,\cite{ReadSachdev_SpN} has been widely used for \su{2} models and more recently for models with \su{4} symmetry\cite{SBMFT_SU4} and \su{N} models.\cite{SBMFT_SUN,SBMFT_SU3}
This approach is justified in the context of a large $\mathcal N$ expansion, where $\mathcal N$ is the number of boson flavors.
Here, we stress that $N$ and $\mathcal N$ are two different numbers and the mean-field approach is equivalent to taking the limit $\mathcal N$ to infinity with $N$ fixed.
After a brief summary of the SBMFT, we address the following question: Is this theory really adapted to the study of models with \su{N} symmetry when $N>2$?

We use the Schwinger bosons defined in Sec.~\ref{sec:LFWT} and impose the constraint on the boson number at each lattice site $i$:
\begin{equation}
 \widehat n(i)=\sum_\alpha b^\dag_\alpha(i) b_\alpha(i)=\kappa,
\label{eq:constraint}
\end{equation}
For the model of Eq.(\ref{eqn:su3}), $\kappa=1$, and the Hamiltonian is the same as Eq.~\ref{eq:H_bosons}.
We define the link operator $\widehat A_{ij\alpha\beta}=(b_\alpha(i)b_\beta(j)-b_\alpha(j)b_\beta(i))/\sqrt 2$ such that the permutation operator writes
\begin{equation}
\widehat P_{ij} =1-\sum_{ \alpha>\beta}2\widehat A^\dag_{ij\alpha\beta}\widehat A_{ij\alpha\beta}.
 \label{eq:Pij}
\end{equation}
The Hamiltonian is thus of degree four in bosonic operators and is not directly solvable.
A mean-field (MF) approximation lowers the degree to two: the Hamiltonian becomes quadratic and solvable by a Bogoliubov transformation.
One possible MF parameter is $\Delta_{ij\alpha\beta}=\langle\widehat A_{ij\alpha\beta}\rangle$.
It is the most often used in \su{2} SBMFT because it is invariant by \su{2} transformations : $\widehat A_{ij\alpha\beta}$ is the destruction operator of a \su{2} singlet of colors $\alpha$ and $\beta$.
The MF Hamiltonian writes
\begin{eqnarray}
 H_{\rm MF}&=&\sum_{\langle ij\rangle }\left(1-2\sum_{\alpha>\beta}(\Delta_{ij\alpha\beta}\widehat A_{ij\alpha\beta}^\dag+h.c.-|\Delta_{ij\alpha\beta}|^2)\right)
 \nonumber\\
 &&+\sum_i\lambda(i)(\kappa-\widehat n(i)),
 \label{eq:Ham_SBMFT}
\end{eqnarray}
where a Lagrange multiplier $\lambda(i)$ is used to impose the constraint of Eq.~\ref{eq:constraint} on average at each site.

The success encountered by this theory for \su{2} models comes from the fact that two phases are possible for the ground state of the Hamiltonian of Eq.~\ref{eq:Ham_SBMFT}.
For $\kappa$ lower than a critical value $\kappa_c$, the ground state is invariant by \su{2} global spin transformations and the elementary excitations are gapped spinons. If the Hamiltonian symmetries are respected,\cite{Wen_PSG,PSG} this phase is a topological spin liquid.
For $\kappa>\kappa_c$, one or several spinons become gapless, and to satisfy the constraint, a Bose condensate is required, breaking the \su{2} symmetry of the ground state: we have a long range ordered state.
Unlike spin-wave expansions, this theory does not assume any order.
The system is free to order or not, and the pattern is not imposed.

This property comes from the fact that the MF Hamiltonian (Eq.~\ref{eq:Ham_SBMFT}) is expressed in terms of \su{2} invariant link operators.
For $N>2$, singlets occupy $N$ sites and can no longer be destroyed by quadratic operators.
This implies that the MF Hamiltonian is no longer \su{N} invariant.
This can be verified using the order parameter $m$ obtained using the Casimir operator on a site
\begin{equation}
\label{eq:m}
 m=\sqrt{\frac{1}{N-1}\left(N\sum_{\alpha,\beta}\langle S_\alpha^\beta\rangle^2-1\right)}.
\end{equation}
For a product (non entangled) state, $m=1$ and for a site in a \su{N} singlet, $m=0$ .
For \su{2}, we recover $m=|\langle \mathbf S\rangle|$.
For $N>2$, the only way to have $m=0$ would be to set all the MF parameters $\Delta_{ij\alpha\beta}$ to 0.
Thus even without any gapless spinon, the ground state is already long range ordered and we cannot have a spin liquid.
The condensation is then only the breaking of a remaining freedom of the bosons.

We now come back to the \su{3} model on the square and triangular lattice.
The $\kappa\to\infty$ limit is the classical limit, namely the $3$ state Potts model.
We first calculate the MF parameters in this limit for the orders considered in this article, with 2 (or 3) sublattices denoted $A$, $B$ (and $C$).
These orders still have the degeneracies associated to the global \su{3} symmetry.
Depending on the choice of 3 orthogonal vectors ${\bf u}_X,\,X=A,B,C$, which specify the orientation on the different sublattices, the mean-field parameters $\Delta_{ij\alpha\beta}$ take different phases and modulus, unlike \su{2} where the only degree of freedom was the gauge choice (the values of $\Delta_{ij\alpha\beta}$ did not depend on how \su{2} was broken).
We once again note that a \su{3} spin liquid cannot be described in this formalism.
Let us choose
\begin{equation}
        {\bf u}_A=\begin{pmatrix}
         1\\0\\0
        \end{pmatrix},
        {\bf u}_B=\begin{pmatrix}
         0\\1\\0
        \end{pmatrix},
        {\bf u}_C=\begin{pmatrix}
         0\\0\\1
        \end{pmatrix}.
        \nonumber
       \end{equation}
In the $\kappa\to\infty$ limit, we can replace the $b_\alpha(i)$ operators by their mean values $\langle b_\alpha(i)\rangle$.
If $i$ is on the $X$ sublattice, $\langle b_\alpha(i)\rangle=({\bf u}_X)_\alpha$.
Thus we obtain the MF parameters for all the considered orders: $\Delta_{ij\alpha\beta}=(\langle b_\alpha(i)\rangle\langle b_\beta(j)\rangle-\langle b_\alpha(j)\rangle\langle b_\beta(i)\rangle)/\sqrt 2$.

\begin{table}
\begin{center}
   \begin{tabular}{|c|c|c|c|c|}
  \hline
  lattice&state&$E_{\rm MF}$&$m$&$n_c$\\
  \hline
triangular&three sublattices&-0.58&0.93&0.86\\
square&two sublattices&-0.68&0.50&0.61\\
square&three sublattices&-0.62&0.89&0.73\\
  \hline
 \end{tabular}
\end{center}
\label{tab:SBMFT_results}
\caption{Results obtained by SBMFT.
$E_{\rm MF}$ is the energy per site obtained by Eq.~\ref{eq:Ham_SBMFT} with mean-field parameters verifying the self-consistency conditions.
$m$ is defined in Eq.~\ref{eq:m} and $n_c$ is the number of bosons in the gapless modes(s).
These quantities are extrapolated to the thermodynamical limit.
}
\end{table}

For finite $\kappa$, these parameters are not self-consistent and have to be adjusted.
The chemical potential $\lambda$ is assumed to be site independant.
We restrict our search for mean-field solutions to states obtained from the classical states by changing the modulus of the non zero $\Delta$'s.
The energy, the order parameter and the fraction of condensed bosons  $n_c$ are given in Tab.~\ref{tab:SBMFT_results} for the states discussed in this article.
In all cases, $m$ is far larger than the order parameter obtained in LFWT, DMRG and iPEPS, even without any boson condensation.
The energies obtained are not variational since the boson Hilbert space is larger than the physical one (the ground state is a superposition of states with different boson numbers on each site).

The three sublattice state on the square lattice has a larger energy than the two sublattice state, in contrast to the results obtained with LFWT, ED, DMRG and iPEPS.
This result can be understood qualitatively.
The \su{3} two-sublattice and the \su{2} two-sublattice state (historically treated by Arovas and Auerbach\cite{Auerbach_largeN}) share several properties: same energy and same condensed fraction.
The value of $m=0.5$ of the order parameter for \su{3} is exactly the value for a site in a \su{2} singlet and corresponds to $m=0$ for \su{2}.
By fixing the MF parameters, a direction among the three available is forbidden to the bosons, that are confined in a \su{2} manifold. This remaining symmetry can only be broken by a condensate.
The \su{3} ground state is thus exactly the same as the \su{2} one.
The only difference stays in the existence of additional excitations in the third direction for \su{3}.
They have a maximal energy cost, the value of the chemical potential, and form a flat band over all other excitations.
Thus, the energy cannot be lowered by fluctuations in the full \su{3} space.
We see in Tab.~\ref{tab:SBMFT_results} that the magnetization is even larger in the other states (the three sublattice states on the square and triangular lattice), with $m=0.89$ and $0.93$ as compared to $m=0.5$.
The fluctuations are even more constrained, which provides a plausible explanation why the energy of the three-sublattice state on the square lattice is larger than the energy of the two sublattice state.

To conclude this appendix, let us put these results in a broader perspective. For SU(2), the
SBMFT is a convenient way to go beyond linear spin-wave theory (LSWT). This for instance allows
to lift the classical degeneracies that may survive LFWT.
The kagome antiferromagnet offers a good example.
Within LSWT, coplanar classical ground states remain degenerated, whereas SBMFT lifts the degeneracy in favor of the $\sqrt{3}\times\sqrt{3}$.\cite{Sachdev}
For that model, a classical state with higher linear spin-wave energy has recently been shown
to have an even lower SBMFT energy.\cite{cuboc1}
In the context of the \su{3} model on the square lattice, the motivation to use SBMFT was to
remove the line of soft modes obtained in LFWT for the three-sublattice order, and at the same
time the divergence of the correction to the magnetization. This is indeed achieved by the SBMFT,
but unfortunately the resulting picture is not satisfactory: For \su{N} models with $N>2$, a \emph{spontaneous} breaking of the \su{N} symmetry is not possible since the choice of MF parameters already breaks it.
No spin liquid ground state exists for the MF Hamiltonian of Eq.~\ref{eq:Ham_SBMFT} and the quantum fluctuations taken into account in a long-range ordered ground-state are limited to some subspace of \su{3}.
As a consequence, we cannot use it to compare the MF states derived from several \su{3} classical orders.
Other bosonic representations of \su{N} spins could lead to better results.\cite{SBMFT_nematic}

\bibliographystyle{apsrev4-1}
\bibliography{refs,SU3_SBMFT}

\end{document}